\documentclass[journal]{IEEEtran}
\usepackage{booktabs}
\usepackage{amsfonts}
\usepackage{amsmath}
\usepackage[style=ieee,maxnames=4,minnames=3,maxbibnames=3]{biblatex}

\usepackage{array}
\usepackage{mdwmath}
\usepackage{mdwtab}
\usepackage{amssymb}
\usepackage{pifont}
\usepackage[utf8]{inputenc}   
\DeclareUnicodeCharacter{202F}{\nobreakspace}
\usepackage{graphicx,epsfig}
\usepackage{eqparbox}
\usepackage{subfigure}
\usepackage{caption}
\usepackage{mathrsfs}
\usepackage{url}
\usepackage{amssymb}
\usepackage{float}
\usepackage{indentfirst}
\usepackage{tabularx,ragged2e}
\usepackage{cases}
\usepackage{algorithm}
\usepackage{algorithmic}
\usepackage{mathtools}
\usepackage{bm}
\usepackage{enumitem}
\usepackage{setspace}
\usepackage{makecell}
\usepackage{xcolor}
\usepackage{ntheorem}
\usepackage{epstopdf}
\usepackage{verbatim}
\usepackage{enumerate}
\usepackage{color}
\usepackage{subfigure}
\theoremheaderfont{\bfseries}
\theorembodyfont{\upshape}
\theoremseparator{:}

\usepackage{multirow}

\newcolumntype{C}{>{\Centering\arraybackslash}X}

\begin{document}  
\title{\Huge Agentic AI-RAN Empowering Synergetic Sensing, Communication, Computing, and Control}
\author{\IEEEauthorblockN{
Lingxiao Sun, 
Zhaoyang Zhang, 
Zihan Lin,  
Zirui Chen,
Weijie Zhou,
Zhaohui Yang,
and Tony Q.~S.~Quek}
\thanks{Lingxiao Sun, Zhaoyang Zhang (corresponding author), Zihan Lin, Zirui Chen, Weijie Zhou, and Zhaohui Yang are with College of Information Science and Electronic Engineering, Zhejiang University, Hangzhou 310027, China, and also with Zhejiang Provincial Key Laboratory of Multi-modal Communication Networks and Intelligent Signal Processing, Hangzhou 310027, China.}
\thanks{T. Q. S. Quek is with the ISTD Pillar, Singapore University of Technology and Design (SUTD), Singapore 487372, and also with the SUTD-ZJU IDEA Center of Network Intelligence, Singapore 487372.}
}
\maketitle 
\begin{abstract}
Future sixth-generation (6G) networks are expected to support low-altitude wireless networks (LAWNs), where unmanned aerial vehicles (UAVs) and aerial robots operate in highly dynamic three-dimensional environments under stringent latency, reliability, and autonomy requirements. In such scenarios, autonomous task execution at the network edge demands holistic coordination among sensing, communication, computing, and control (SC3) processes. Agentic Artificially Intelligent Radio Access Networks (Agentic AI-RAN) offer a promising paradigm by enabling the edge network to function as an autonomous decision-making entity for low-altitude agents with limited onboard resources. 
In this article, we propose a task-oriented Agentic AI-RAN architecture that enables SC3 task execution within a single edge node. The proposed architecture addresses the challenge of coordinating heterogeneous workloads in resource-constrained edge environments. To validate this framework, we prototype a representative low-altitude UAV system on a general-purpose Graphics Processing Unit (GPU) platform and evaluate it through an autonomous drone-navigation case study. The current prototype instantiates the platform-agnostic design through Multi-Instance GPU (MIG) partitioning and containerized deployment, providing physical resource isolation and coordinated execution between real-time communication and multimodal inference. Experimental results demonstrate low closed-loop latency, robust bidirectional communication, and stable performance under dynamic runtime conditions, highlighting the feasibility of the proposed framework for mission-critical low-altitude wireless networks in 6G.

    \end{abstract}

\IEEEpeerreviewmaketitle

\section{Introduction}
The sixth-generation (6G) wireless networks are envisioned to deliver not only ultra-high-speed connectivity but also deterministic performance to support autonomous, real-time operations. This evolution is particularly critical for emerging low-altitude wireless networks, where unmanned aerial vehicles (UAVs) and aerial robots must operate in complex, three-dimensional environments under stringent latency, reliability, and autonomy constraints \cite{song2021survey}. Unlike terrestrial devices, low-altitude platforms face severe Size, Weight, and Power (SWaP) limitations that restrict onboard computing capabilities. As a result, they can neither rely solely on local processing for complex missions nor depend on centralized cloud control due to unstable air-to-ground links. These constraints motivate a shift toward edge-enabled autonomous control, in which the network edge functions as an external decision logic entity located close to data sources and actuators. Such an external brain must simultaneously process high-bandwidth sensory data and sustain reliable control links, placing significant and heterogeneous demands on edge computing resources.

Recent advances in agentic artificial intelligence (Agentic AI) provide a promising foundation for addressing this challenge. This paradigm reconceptualizes edge nodes not merely as data relays, but as intelligent agents capable of autonomously executing complete perception\textendash{}reasoning\textendash{}control cycles \cite{11370176}. Within this context, Agentic AI-RAN emerges as a unifying framework that orchestrates sensing, communication, computation, and control (SC3) as a single closed-loop task. Through an orchestration layer, the network can support context-aware, responsive task execution in real time, which are essential for mission-critical applications in low-altitude environments. Crucially, this agentic execution model shifts intelligence from offline optimization or supervisory control toward continuous interaction guided by feedback from both the physical environment and the wireless system state. This integration demands that a single edge node jointly support semantic reasoning and timing-critical control within a unified execution loop.

\begin{figure*}[t] 
  \centering
  \includegraphics[width=1\linewidth]{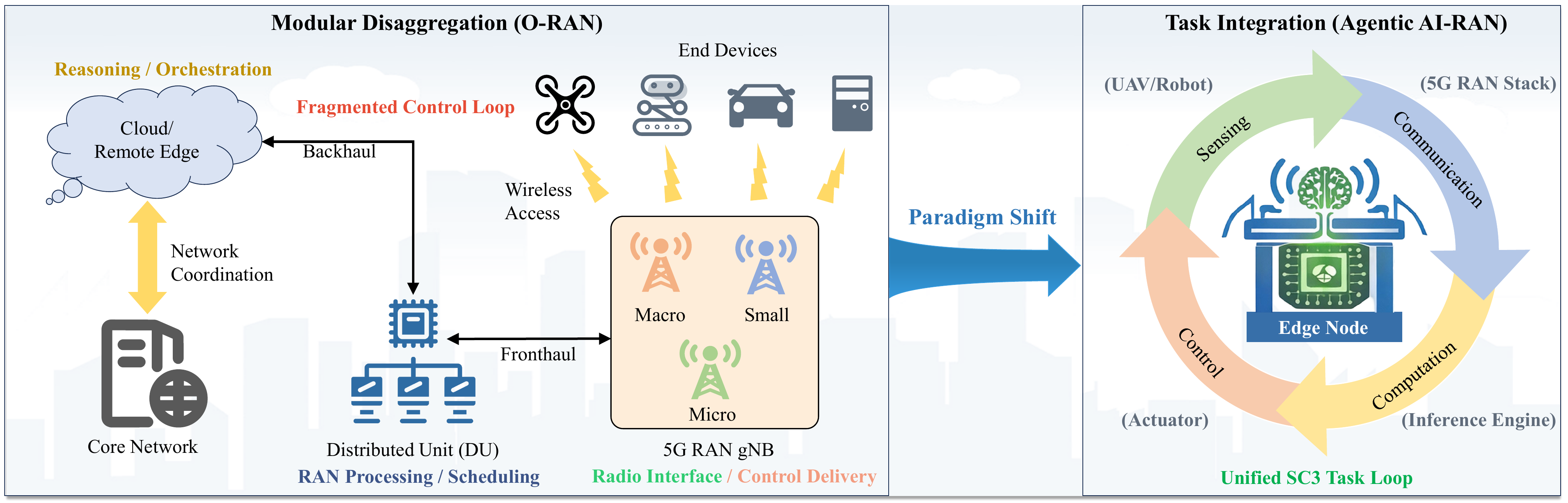}
  \captionsetup{justification=raggedright,singlelinecheck=false}
  \caption{From fragmented processing to edge-native autonomous control.}
  \label{compare}
\end{figure*}

Realizing tightly coupled autonomy within existing network architectures reveals a fundamental paradigm conflict, as illustrated in Fig.~\ref{compare}. The evolution of Radio Access Networks (RAN), exemplified by Open RAN (O-RAN), has improved flexibility through modular disaggregation and multi-vendor interoperability. However, this design orientation toward logical and physical separation is not well aligned with agentic SC3 workloads that require tightly coordinated, low-latency closed-loop execution. A deeper challenge lies in the heterogeneous runtime characteristics of SC3 processing: communication stacks require strict determinism and fine-grained scheduling, whereas AI inference is bursty, memory-intensive, and compute-intensive \cite{10579546}. When these contrasting workloads share hardware resources without effective coordination and isolation, bursty AI execution can undermine communication and control functions that depend on strict timing, leading to jitter, delayed actuation, and reduced task reliability. These observations suggest that SC3-oriented Agentic AI cannot be supported by merely colocating individual functions on a shared platform. Instead, it requires an integrated system design in which perception, reasoning, communication, and actuation operate as a coordinated closed loop under practical latency and resource constraints \cite{10849561}.

To address these challenges, this article proposes a task-oriented Agentic AI-RAN architecture as a general system and workflow framework for SC3 task execution at the network edge. The architecture unifies task-level orchestration with bounded execution domains for heterogeneous SC3 workloads, enabling stable closed-loop execution under practical latency and resource constraints. Within this framework, hardware partitioning, isolated software runtimes, and reasoning-driven task planning provide concrete means to support coordinated SC3 execution on shared edge platforms. The key contribution lies in the vertical integration of bounded execution domains, software isolation, and task-level reasoning to support stable SC3 closed-loop execution on shared edge platforms. A representative prototype is instantiated through Multi-Instance GPU (MIG) partitioning and containerized deployment to validate the framework in an autonomous drone navigation scenario. The rest of this article presents representative SC3 application scenarios, introduces the proposed architecture and execution workflow, and validates the framework through a prototype for autonomous drone navigation, followed by a discussion of future research directions for edge intelligence in 6G networks.

\begin{figure*}[t]
    \centering
    \includegraphics[width=1\linewidth]{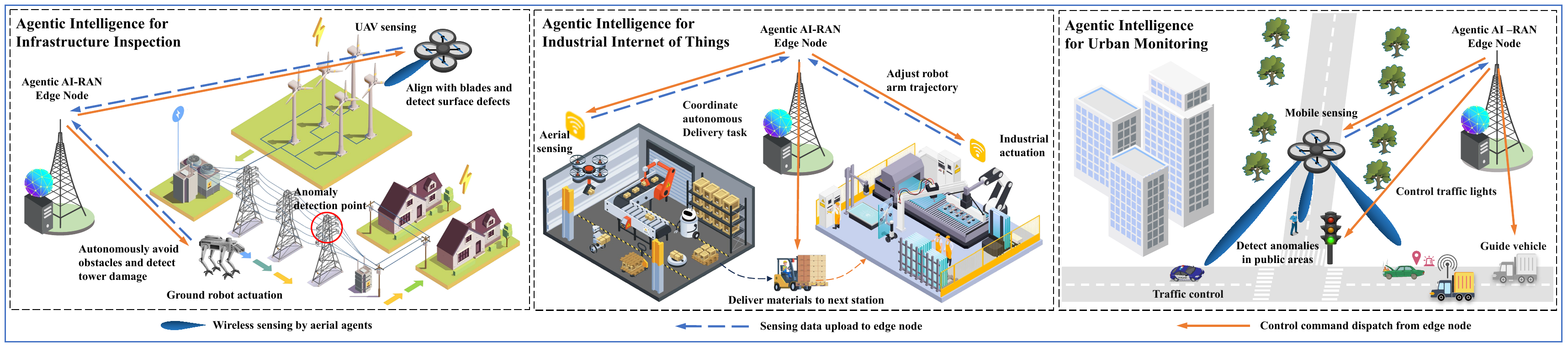}
    \captionsetup{justification=raggedright,singlelinecheck=false}
    \caption{Representative Agentic AI-RAN scenarios featuring SC3 functionality in low-altitude wireless networks.}
    \label{application}
\end{figure*}

\section{Typical Low-Altitude Applications of Agentic AI-RAN}
Agentic AI-RAN extends the role of the network edge beyond connectivity provisioning, positioning it as an edge-resident execution substrate for SC3 tasks. In practical deployments, these edge nodes are tightly coupled with heterogeneous low-altitude agents, including UAVs, aerial robots, and ground robotic systems operating under aerial supervision \cite{jin2025advancing}. In this setting, embedded intelligence is supported by the close integration of edge infrastructure, communication support, and task-level execution for low-altitude robotic applications. These platforms must function in complex three-dimensional environments characterized by dynamic wireless conditions, fluctuating computational demands, and stringent safety requirements. In the scenarios of Fig. \ref{application}, an edge orchestration layer operates close to the radio stack, while the underlying 5G infrastructure provides transport and bounded execution support for perception and control. By consolidating multimodal perception, semantic interpretation, and closed-loop coordination within this edge framework, Agentic AI-RAN enables localized decision making and timely actuation for low-altitude agents under practical resource constraints.

\subsection{Autonomous Infrastructure Inspection}
Low-altitude robotic platforms enable autonomous inspection of critical infrastructure in scenarios where human access is costly, hazardous, or inefficient. Assets such as wind turbines, transmission towers, and power lines require aerial platforms to operate in close proximity to large-scale structures, often under strong electromagnetic interference and complex airflow conditions. As depicted in Fig.~\ref{application}, the edge orchestration framework supports a heterogeneous team comprising low-altitude UAVs and ground robots. While the UAV autonomously aligns with rotating blades to detect surface defects from the air, the robot dog navigates rugged terrain to identify tower damage at ground level. The unified SC3 execution loop enables rapid anomaly detection and synchronized air-ground task execution through timely sensing-data transport and control delivery, significantly improving inspection efficiency and operational safety compared to traditional manual teleoperation.

\subsection{Industrial Internet of Things}
The Industrial Internet of Things (IIoT) is evolving toward highly autonomous production environments where intelligence extends beyond the factory floor into the aerial domain. In large industrial facilities and logistics hubs, low-altitude UAVs increasingly complement ground-based Automated Guided Vehicles (AGVs) and robotic arms to enable three-dimensional situational awareness and flexible material handling \cite{10926519}. Agentic AI-RAN facilitates air-ground coordination by enabling the edge node to host task-level services and fuse state information from heterogeneous agents. For example, UAVs can assist AGVs in navigation and inventory verification from above, while the orchestration layer synchronizes their trajectories through sensing-data transport and control delivery to prevent collisions and optimize task allocation. Upon detecting production deviations through real-time visual inspection, the edge platform can support application-side control services that update robotic arm trajectories without halting the assembly line. This enables precise material handover and continuous operation under tight latency constraints.

\subsection{Urban Monitoring}
With increasing urban density and the emergence of low-altitude aerial mobility, maintaining real-time situational awareness has become critical to public safety and traffic efficiency. Conventional urban surveillance systems rely primarily on fixed ground-based sensors, offering limited spatial coverage and delayed response. In contrast, Agentic AI-RAN enables low-altitude aerial sensing by integrating UAVs as mobile perception nodes that provide a complementary top-down perspective unavailable to terrestrial sensors. By fusing data from patrol drones, roadside units, and traffic infrastructure, the edge platform can host anomaly-detection services for congestion, accidents, or crowd aggregation in public areas. Based on this timely insight, the orchestration layer can support application-level actions such as traffic light adjustment for emergency routing or vehicle guidance toward alternative paths. In this way, urban monitoring evolves from passive observation toward more proactive and decentralized response under tight end-to-end latency constraints.

\begin{figure*}[t] 
    \centering
    \includegraphics[width=1\linewidth]{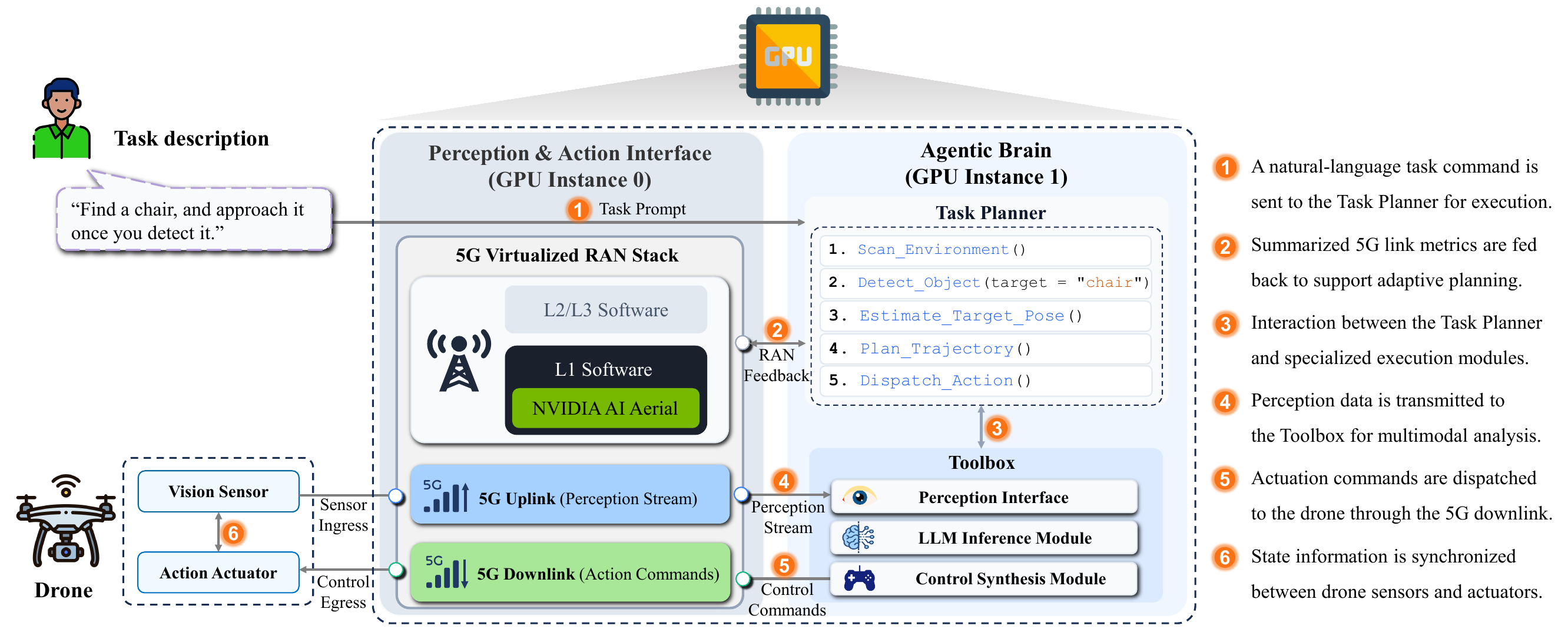}
        \captionsetup{justification=raggedright,singlelinecheck=false}
    \caption{ SC3 task execution flow under bounded communication and inference execution domains.}
    \label{model}
\end{figure*}

\section{Fundamental Challenges in Supporting Agentic AI-RAN}
Agentic AI-RAN faces distinct challenges hindering reliable SC3 execution. Unlike throughput-oriented terrestrial networks, it must support interdependent tasks under strict safety constraints \cite{huang2024communication}. In  dynamic aerial environments, fluctuating channel conditions and bursty inference demands introduce substantial runtime variability, increasing the risk of resource contention and execution jitter. For UAVs relying on fast closed-loop control, such instability can directly compromise flight safety \cite{11141658}. Addressing these challenges necessitates fine-grained coordination across domains together with robust isolation between computing and communication resources. The following subsections highlight two core limitations that hinder reliable agentic execution.

\subsection{Resource Contention from Task Heterogeneity}
SC3 execution at the edge requires coordinating tightly coupled workloads that nevertheless exhibit very different runtime characteristics under shared resource constraints. Communication functions, such as 5G RAN protocol processing, demand deterministic execution with stringent timing requirements on real-time stacks, while AI inference, particularly large language models (LLMs), is bursty and requires substantial memory resources. In conventional edge servers, these workloads contend for shared resources like memory bandwidth and GPU queues, lacking fine-grained isolation \cite{kundu2023hardware}. For instance, a surge in visual inference triggered by complex scene analysis can monopolize memory bandwidth, indirectly starving the radio protocol stack. In low-altitude wireless networks, where edge nodes must simultaneously support perception informed inference and control functions for aerial platforms, such interference manifests not only as increased latency but also as control jitter or missed actuation deadlines, disrupting closed-loop stability and potentially leading to mission failure. This class of interference highlights the need for workload-aware resource partitioning and stronger isolation across both hardware and software layers.

\subsection{Lack of Cross-Domain Scheduling}
The fragmentation of execution environments across communication and AI domains represents another fundamental limitation. Radio processing with strict latency requirements typically runs on specialized networking stacks such as the Data Plane Development Kit (DPDK) on real-time operating systems, whereas AI inference executes on accelerator-based runtimes such as the Compute Unified Device Architecture (CUDA), optimized for throughput and batch processing. These heterogeneous runtimes follow different scheduling semantics, preventing unified visibility into task criticality across sensing, communication, and control. As a result, existing orchestration mechanisms at the resource or container level cannot support semantic-aware preemption or real-time arbitration across SC3 tasks. Since edge nodes are expected to sustain stable SC3 loops under strict safety requirements in low-altitude wireless networks, the lack of a unified runtime abstraction prevents consistent prioritization of sensing and control tasks over best effort background workloads, thereby undermining reliable closed-loop execution \cite{11199308}.

\section{Architecture Design for SC3 Integration in Low-Altitude Wireless Networks}
In low-altitude wireless networks, achieving closed-loop SC3 execution in Agentic AI-RAN requires more than simple resource partitioning. It requires an edge architecture that jointly supports sensing, communication, computing, and control under shared latency and resource constraints. Communication and control functions require deterministic execution, whereas multimodal inference is bursty and resource intensive \cite{11141658}. Without clear execution boundaries, contention in shared resources can directly degrade control fidelity and system responsiveness. To address the challenges identified in Section III, the proposed architecture combines task-level orchestration with vertically integrated isolation, enabling tightly coupled SC3 execution within a single edge node.

\subsection{The Agentic Brain: Unified Task Orchestration}
The core of the proposed architecture is the Agentic Brain, a task-orchestration layer that comprises the Task Planner, Contextual Memory, and Toolbox. As shown in Fig. ~\ref{model}, the Agentic Brain translates high-level mission intents into executable SC3 actions by jointly considering perception outcomes, communication status, and control requirements. With support from a vision-language model (VLM) for semantic interpretation, this orchestration layer coordinates task planning, state tracking, and tool invocation within a unified execution framework.

\textit{1) Task Planner:}
The Task Planner interprets high-level task instructions and converts them into an ordered sequence of executable SC3 actions. Specifically, the planner maps semantic intent to concrete operations such as environment scanning, target detection, trajectory generation, and action dispatch. When task complexity increases, the planner further decomposes the mission into multiple coordinated steps, thereby enabling coherent closed-loop execution under changing runtime conditions.

\textit{2) Contextual Memory:}
To support consistent decisions across the SC3 loop, the Agentic Brain maintains a lightweight contextual memory operating at the task level. The memory records recent perception outcomes, executed actions, task progress, and communication status indicators exported by the 5G RAN stack. Instead of synchronously accessing the fine-grained internal state of the real-time communication scheduler, the orchestration layer relies on asynchronously updated coarse-grained key performance indicator (KPI) exports and summarized runtime snapshots. These task-timescale snapshots allow the system to remain aware of current network conditions while preserving the deterministic execution path of radio processing.

\textit{3) Toolbox Interface:}
SC3 capabilities are exposed through a unified toolbox of callable functions. Core 5G RAN functions, media-processing functions, and drone control primitives are wrapped as standardized interfaces for invocation during task execution. This interface abstraction separates task-level orchestration from low-level execution, while preserving deterministic boundaries for communication and control operations. Under this design, different technical capabilities can be invoked on demand according to task requirements, and new task-specific tools can be integrated through standardized interfaces without redefining the underlying SC3 execution architecture. The present prototype focuses on single-edge-node execution, while cross-node orchestration is outside the current scope.

\begin{table*}[t]
  \centering
  \renewcommand{\arraystretch}{1.8}
  \setlength{\extrarowheight}{2pt}
  \caption{Comparison of AI-RAN Architectures: From Functional Augmentation to Agentic SC3 Integration.}
  \label{table1}
  \resizebox{\textwidth}{!}{%
  \begin{tabular}{c|c|c|c|c}
    \hline\hline
    \textbf{Architecture Type} &
    \textbf{SC3 Loop Completeness} &
    \textbf{Task Planning Mode} &
    \textbf{Hardware Isolation} &
    \textbf{Deterministic Control} \\
    \hline\hline
    Sensing--Communication Fusion \cite{liu2022integrated} & Partial (Sensing + Comm) & N/A & N/A & $\times$ \\
    \hline
    Conventional MEC \cite{mach2017mobile} & Partial (Comp + Comm) & Static Offloading & Software-only & $\times$ \\
    \hline
    Networked Edge Control Systems \cite{khan2019edge} & Modularly Disjoint & Rule-based & Partial & $\checkmark$ \\
    \hline
    \textbf{Proposed Agentic AI-RAN} & \textbf{Integrated} & \textbf{Semantic Planning} & \textbf{Hardware} & $\checkmark$ \\
    \hline\hline
  \end{tabular}
  }
\end{table*}

\subsection{Vertical Integration of Partitioning and Isolation}
The proposed architecture adopts a vertically integrated design that combines hardware-level resource partitioning with software-level runtime isolation. Its objective is to establish clear execution boundaries between heterogeneous SC3 workloads with substantially different timing requirements and resource demands. At the architectural level, communication and control functions operate within predictable execution domains, while perception and multimodal inference are confined to separate domains so that bursty inference does not interfere with timing-critical tasks. Such separation can be realized in heterogeneous computing environments through dedicated accelerator slices, isolated processing domains, hardware virtualization, or other forms of resource reservation. Although the specific mechanism may differ across platforms, the architectural requirement is unchanged: compute-intensive inference and real-time network execution must not contend in an uncontrolled manner for shared resources. In practice, this design requires three basic capabilities: enforceable separation between real-time communication and bursty inference workloads, isolated software runtimes for major functional components, and a controlled interface through which the orchestration layer can access summarized runtime feedback without entering the timing-critical execution path. On top of this substrate, logical functions can be mapped to bounded execution domains in different hardware and software forms while preserving the same SC3 execution logic. MIG partitioning and containerized deployment are therefore concrete realizations of this broader design.

\subsection{The Agentic Workflow: From Perception to Action}
Building upon the Agentic Brain and vertical integration, the proposed architecture realizes a continuous SC3 closed loop \cite{10926519}. The workflow begins with multimodal perception, where heterogeneous inputs including video streams, task prompts, and wireless context information from the 5G RAN stack are fused into a structured execution context. Based on this context, the Agentic Brain performs task-aware reasoning and planning. A locally deployed VLM interprets semantic intent, evaluates situational cues, and decomposes high-level objectives into executable control strategies by jointly reasoning over task objectives, perception outcomes, and communication constraints. These strategies are then executed through callable tools. Through the Toolbox abstraction, control actions are dispatched via the 5G downlink while communication and runtime-level adjustments are applied within the virtualized RAN and edge runtime. Execution outcomes are continuously fed back to update the execution state for subsequent decisions. Through this tightly coupled perception, reasoning, and action cycle, the system regulates the SC3 process toward mission objectives and enables robust autonomous operation in Agentic AI RAN deployments.

\begin{figure*}[t] 
    \centering
    \includegraphics[width=1\linewidth]{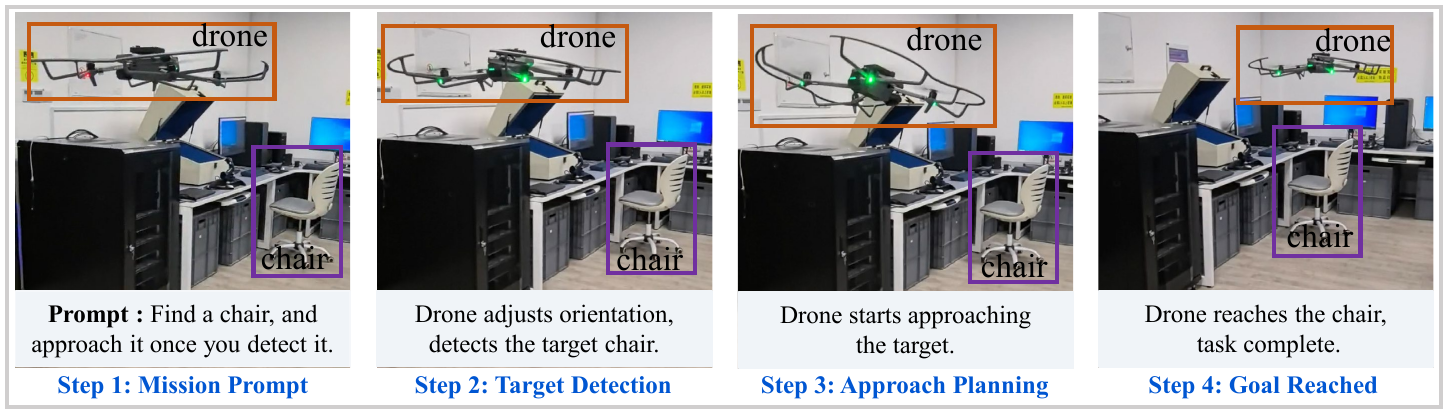}
        \captionsetup{justification=raggedright,singlelinecheck=false}
    \caption{End-to-end task execution of an autonomous drone in an indoor environment.}
    \label{simulation}
\end{figure*}

\begin{figure*}[t] 
    \centering
    \includegraphics[width=1\linewidth]{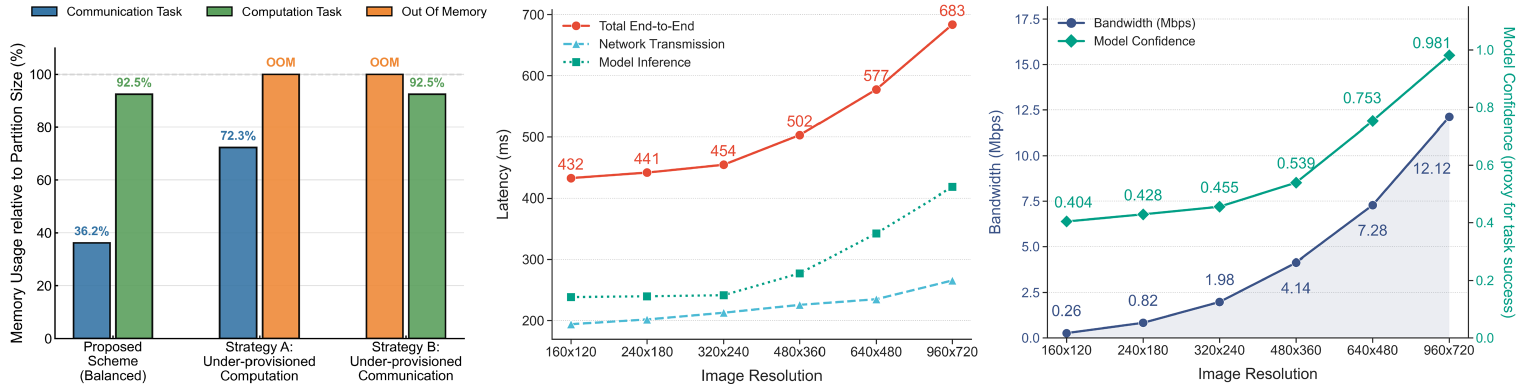}
        \captionsetup{justification=raggedright,singlelinecheck=false}
    \caption{Experimental evaluation of closed-loop SC3 execution in the proposed Agentic AI-RAN system.}
    \label{result}
\end{figure*}

\subsection{Comparison with Other Related Methods}
To better position the proposed architecture within the existing ecosystem, Table \ref{table1} provides a qualitative positioning of the proposed architecture relative to several representative paradigms. Traditional works in sensing-communication fusion and conventional MEC mainly strengthen data relaying and static task offloading \cite{liu2022integrated,mach2017mobile}, whereas modular edge control systems provide deterministic support for control tasks. Building upon these advancements, the proposed Agentic AI-RAN further introduces task-level semantic planning together with vertically integrated hardware-software isolation, allowing high-level instructions to be translated into executable SC3 actions while limiting interference between compute-intensive inference and time-critical RAN functions. Table \ref{table1} positions the proposed framework at the architectural level, while Section V evaluates its behavior using the closest available baseline and under different resource configurations. Rather than replacing existing methods, this architecture is intended to provide a specialized, scalable foundation for mission-critical edge applications where tightly coupled SC3 execution is essential \cite{huang2024communication}.

\section{A Case Study: Closed-Loop Task Execution}
In this section, we present a case study to evaluate the proposed Agentic AI-RAN system for closed-loop SC3 task execution in low-altitude wireless networks. The scenario involves an autonomous drone executing a high-level task based on natural language instructions. The prototype serves as a representative realization of the proposed framework and is deployed on an NVIDIA A100X GPU, where MIG is used to instantiate bounded execution domains for communication and multimodal inference. The UAV platform is a DJI Mavic 3E. The wireless system operates in the 3.5 GHz band with 100 MHz instantaneous bandwidth and a 4T4R antenna configuration. This indoor setup is used to evaluate closed-loop SC3 task execution. A run is counted as successful if the drone completes the instructed task within a prescribed time budget of 15 s without communication failure or control interruption. For each resolution setting, the experiment is repeated 240 cycles, and the reported latency and model confidence values are averaged over these runs unless otherwise stated.

The end-to-end closed-loop task execution process is illustrated in Fig.~\ref{simulation}. Upon receiving a high-level instruction (e.g., ``\textbf{Find a chair and approach it once you detect it.}''), the edge node fuses task intent with visual input and performs semantic reasoning using locally deployed 2.8B DeepSeek-VL model. The generated plan is then executed through instruction parsing, target detection, approach planning, and low-level control delivery, thereby forming a complete SC3 closed loop at the edge. For comparison, we also consider a software-isolated deployment that preserves containerized execution but removes hardware-bounded resource partitioning. Although the functional pipeline remains unchanged, all modules contend for shared GPU copy and compute resources. Under this setting, bursty inference interferes with timing-critical RAN execution and triggers RAN H2D timeout, preventing stable task completion. This result indicates that software-level isolation alone is insufficient for reliable SC3 execution in the present prototype.

We then examine resource dimensioning within the proposed architecture. GPU profiling shows that the communication module consumes approximately 14.5 GB of memory, while the inference module requires about 37.0 GB. As shown in Fig.~\ref{result}, allocating 60\% of resources to communication and 40\% to inference achieves stable operation. By contrast, Strategy A, with only 25\% allocated to computation, causes Out-of-Memory (OOM) failure in the inference engine, while Strategy B, with only 25\% allocated to communication, leads to partition exhaustion, severe link instability, and packet loss. Together, these results characterize the sensitivity of the proposed framework to resource imbalance and show that stable SC3 execution depends on workload-aware partitioning.

Under the stable configuration, the reported latency values are average end-to-end delays measured from the arrival of the task input at the edge node to the delivery of the control command for actuation. As resolution increases from low to high, the total closed-loop latency increases from sub-500 ms to approximately 680 ms, with multimodal inference dominating the delay. Network transmission latency grows more gradually and remains bounded during concurrent inference and control execution under hardware isolation. On the current platform, the end-to-end response is therefore limited mainly by model-side processing rather than communication, while further latency reduction is expected to come from lighter model deployment and compute-aware model scaling.

Finally, we investigate the impact of communication constraints on perception-driven control. Fig.~\ref{result} reports bandwidth consumption and model confidence under different resolution settings, where varying the resolution emulates communication loads from sub-Mbps to over 10~Mbps. Model confidence is taken from the output score of the deployed vision-language model during semantic inference about the target and is used here as an indicator of semantic reliability of perception. Latency and confidence are averaged over repeated runs at each resolution, whereas system-level closed-loop reliability is reflected by a Task Success Rate of about 70\% over 20 repeated trials with varying initial drone positions. Together, these results  reveal a clear trade-off between communication load and perception quality. While the present section focuses on a representative UAV case study, the same SC3 execution logic has also been instantiated in a quadruped-based 5G edge prototype, further indicating that the proposed architecture can extend to heterogeneous robotic agents.

\section{Future Directions}

\subsection{Hardware Heterogeneity and Energy Efficiency}
Although the current prototype leverages high-performance NVIDIA A100X GPUs to validate the feasibility of Agentic AI-RAN, the proposed architecture is inherently agnostic to specific hardware tiers \cite{chen2025towards}. For practical 6G edge deployments, the framework can be adapted to lower-power embedded platforms, such as NVIDIA Jetson Orin or specialized AI-on-5G SoCs. With lightweight virtualization and model compression, SC3 orchestration can also be supported on resource-constrained devices, enabling a better balance between intelligence capability and energy efficiency in large-scale field deployment. Future research should further investigate compute-aware model scaling and latency-energy trade-offs, enabling adaptive inference fidelity under strict latency constraints.

\subsection{Coordinated Resource Management in Multi-Agent Systems}
As Agentic AI-RAN expands to support multiple agents and heterogeneous applications, cross-service resource coordination becomes increasingly important \cite{9839628}. Different tasks may invoke different model stacks, perception pipelines, and control modules through the Toolbox, creating mixtures of control-critical and compute-intensive workloads that compete for shared compute, memory, and network resources. While static partitioning can bound interference within a single edge node, larger-scale deployments require more adaptive mechanisms for allocating resources across concurrent services. Future work should therefore investigate runtime frameworks for context-aware resource arbitration, task-aware model placement, and service-level prioritization, so as to reduce inference latency and scheduling overhead while preserving real-time control guarantees. Large-scale deployment will also require an upper-layer service orchestration function above the local Agentic Brain to support runtime activation, task placement, lifecycle management, and cross-node coordination across concurrent SC3 services.

\subsection{Enabling Distributed Agentic Intelligence}
The current framework supports edge autonomy by enabling individual agents to perform SC3 tasks within a single edge node. However, many real-world applications, such as search and rescue or distributed monitoring, require collaboration among spatially distributed agents. These agents must share contextual knowledge, align task objectives, and  synchronize actions decentrally. A key challenge lies in preventing communication latency amplification and synchronization overhead from undermining real-time SC3 execution. Future architectures should therefore support federated inference, decentralized semantic reasoning, and collaborative model execution. For example, decentralized knowledge exchange and peer-to-peer learning can enable agents to share information and coordinate planning without excessively increasing latency. Realizing such distributed agentic intelligence will require lightweight coordination protocols, robust synchronization mechanisms, and scalable lifecycle management for heterogeneous task-specific runtimes.

\section{Conclusion}
This article proposed an Agentic AI-RAN architecture for SC3 task execution in 6G low-altitude wireless networks. The framework is instantiated through MIG-based partitioning and containerized deployment to enable execution of communication and inference tasks within a single edge node. Experimental results demonstrate that, under hardware-isolated and resource-constrained settings, the system sustains closed-loop latency ranging from 500~ms to 680~ms across visual resolutions, while maintaining robust bidirectional communication. The architecture further provides a scalable design reference for low-altitude applications such as infrastructure inspection, IIoT, and urban monitoring in agentic 6G networks.

\section{Acknowledgment}
This work was supported in part by National Natural Science Foundation of China under Grant 62394292 and the Fundamental Research Funds for the Central Universities under Grant 226-2024-00069.

\printbibliography

@ARTICLE{11141658,
  author={Kundu, Lopamudra and Lin, Xingqin and Gadiyar, Rajesh and Lacasse, Jean-Francois and Chowdhury, Shuvo},
  journal={IEEE Communications Magazine}, 
  title={{AI}-{RAN}: {T}ransforming {RAN} with {AI}-driven {C}omputing {I}nfrastructure}, 
  year={2026},
  volume={64},
  number={1},
  pages={168-174},
}

@ARTICLE{10926519,
  author={Lei, Chengleyang and Fang, Xinran and Feng, Wei and Chen, Yunfei and Xiao, Ming and Ge, Ning and Mao, Shiwen},
  journal={IEEE Network}, 
  title={{S}atellite-{UAV} {N}etworks for {6G} {C}ontrol: {A} {S}ensing-{C}ommunication-{C}omputing-{C}ontrol {C}losed {L}oop {P}erspective},
  year={2025},
  volume={39},
  number={4},
  pages={62-69},
}

@article{huang2024communication,
  title={{C}ommunication and {C}omputing {I}ntegrated {RAN}: {A} {N}ew {P}aradigm {S}hift for {M}obile {N}etwork},
  author={Huang, Yuhong and Li, Nan and Sun, Qi and Li, Xiang and Huang, Jinri and Chen, Ziqi and Xu, Xiaofei and Chih-Lin, I},
  journal={IEEE Network},
  volume={38},
  number={2},
  pages={97--112},
  year={2024},
  publisher={IEEE}
}

@ARTICLE{10579546,
  author={Chen, Zirui and Zhang, Zhaoyang and Yang, Zhaohui},
  journal={IEEE Wireless Communications}, 
  title={{B}ig {AI} {M}odels for {6G} {W}ireless {N}etworks: {O}pportunities, {C}hallenges, and {R}esearch {D}irections},
  year={2024},
  volume={31},
  number={5},
  pages={164-172},
}

@ARTICLE{10849561,
  author={Acharya, Deepak Bhaskar and Kuppan, Karthigeyan and Divya, B.},
  journal={IEEE Access}, 
  title={{A}gentic {AI}: {A}utonomous {I}ntelligence for {C}omplex {G}oals--{A} {C}omprehensive {S}urvey}, 
  year={2025},
  volume={13},
  number={},
  pages={18912-18936},
}

@article{kundu2023hardware,
  title={{H}ardware {A}cceleration for {O}pen {R}adio {A}ccess {N}etworks: {A} {C}ontemporary {O}verview},
  author={Kundu, Lopamudra and Lin, Xingqin and Agostini, Elena and Ditya, Vikrama and Martin, Tim},
  journal={IEEE Communications Magazine},
  volume={62},
  number={9},
  pages={160--167},
  year={2023},
  publisher={IEEE}
}

@ARTICLE{11370176,
  author={Jiang, Feibo and Pan, Cunhua and Wang, Kezhi and Michiardi, Pietro and Dobre, Octavia A. and Debbah, Merouane},
  journal={IEEE J. Sel. Areas Commun.}, 
  title={From {L}arge {AI} {M}odels to {A}gentic {AI}: {A} {T}utorial on {F}uture {I}ntelligent {C}ommunications}, 
  year={2026},
  volume={44},
  number={},
  pages={3507-3540},
}

@article{chen2025towards,
  title={{T}owards {W}ireless {N}ative {B}ig {AI} {M}odel: {T}he {M}ission and {A}pproach {D}iffer from {L}arge {L}anguage {M}odel},
  author={Chen, Zirui and Zhang, Zhaoyang and Liu, Chenyu and Xing, Ziqing},
  journal={Science China Information Sciences},
  volume={68},
  number={7},
  pages={170303},
  year={2025},
  publisher={Springer}
}

@article{jin2025advancing,
  title={{A}dvancing the {C}ontrol of {L}ow-{A}ltitude {W}ireless {N}etworks: {A}rchitecture, {D}esign {P}rinciples, and {F}uture {D}irections},
  author={Jin, Haijia and Yuan, Weijie and Wu, Jun and Wang, Jiacheng and Niyato, Dusit and Wang, Xianbin and Karagiannidis, George K and Lin, Zhiyun and Gong, Yi and Kim, Dong In and others},
  journal={npj Wireless Technology},
  volume={2},
  number={1},
  pages={2},
  year={2026},
  publisher={Nature Publishing Group UK London}
}

@ARTICLE{11199308,
  author={Li, Jingli and Ma, Yiyan and Ai, Bo and Yuan, Weijie and Cheng, Qingqing and Ma, Guoyu and Yang, Mi and Lu, Yunlong and Yue, Wenwei and Zhong, Zhangdui},
  journal={IEEE Transactions on Vehicular Technology}, 
  title={{S}ensing-{E}nhanced {H}andover {C}riterion for {L}ow-{A}ltitude {W}ireless {N}etworks ({LAWNs})}, 
  year={2025},
  volume={},
  number={},
  pages={1-6},
}

@article{liu2022integrated,
  title={{I}ntegrated {S}ensing and {C}ommunications: {T}oward {D}ual-{F}unctional {W}ireless {N}etworks for {6G} and {B}eyond},
  author={Liu, Fan and Cui, Yuanhao and Masouros, Christos and Xu, Jie and Han, Tony Xiao and Eldar, Yonina C and Buzzi, Stefano},
  journal={IEEE Journal on Selected Areas in Communications},
  volume={40},
  number={6},
  pages={1728--1767},
  year={2022},
  publisher={IEEE}
}

@article{mach2017mobile,
  title={{M}obile {E}dge {C}omputing: {A} {S}urvey on {A}rchitecture and {C}omputation {O}ffloading},
  author={Mach, Pavel and Becvar, Zdenek},
  journal={IEEE Commun. Surveys Tuts.},
  volume={19},
  number={3},
  pages={1628--1656},
  year={2017},
  publisher={IEEE}
}

@article{khan2019edge,
  title={{E}dge {C}omputing: {A} {S}urvey},
  author={Khan, Wazir Zada and Ahmed, Ejaz and Hakak, Saqib and Yaqoob, Ibrar and Ahmed, Arif},
  journal={Future Generation Computer Systems},
  volume={97},
  pages={219--235},
  year={2019},
  publisher={Elsevier}
}

@article{song2021survey,
  title={{A} {S}urvey of {P}rototype and {E}xperiment for {UAV} {C}ommunications},
  author={Song, Qingheng and Zeng, Yong and Xu, Jie and Jin, Shi},
  journal={Science China Information Sciences},
  volume={64},
  number={4},
  pages={140301},
  year={2021},
}

@ARTICLE{9839628,
  author={Abdalla, Aly S. and Upadhyaya, Pratheek S. and Shah, Vijay K. and Marojevic, Vuk},
  journal={IEEE Netw.}, 
  title={Toward Next Generation Open Radio Access Networks: What O-RAN Can and Cannot Do!}, 
  year={2022},
  volume={36},
  number={6},
  pages={206-213},
}

\end{document}